\documentstyle[12pt,psfig] {article}
\def\CSV{{\scriptscriptstyle{CSV}}}
\def\CSB{{\scriptscriptstyle{CSB}}}
\def\EM{{\scriptscriptstyle{EM}}}
\begin{document}
\thispagestyle{empty}
\title{\flushright{IU/NTC 98-01}
 \center{Flavor and Charge Symmetry in the Parton Distributions of 
 the Nucleon}}
\author{ C.J. Benesh$^*$ and J.T. Londergan,\\
Nuclear Theory Center, Indiana University,\\
Bloomington, Indiana 47405\\
$^*$ Current Address: Wesleyan College, 4760 Forsyth Road, Macon, Ga. 31204}
\maketitle
\setcounter{page}{0}

\begin{abstract}

    Recent calculations of charge symmetry violation (CSV) in the valence
quark distributions of the nucleon have revealed that the dominant
symmetry breaking contribution comes from the mass associated with the 
spectator quark system. Assuming that the change in the spectator mass
can be treated perturbatively, we derive a model independent expression
for the shift in the parton distributions of the nucleon. This result is 
used to derive a relation between the charge and flavor 
asymmetric contributions to the valence quark distributions in the proton,
and to calculate CSV contribution to the nucleon sea. The CSV contribution
 to the Gottfried sum rule is also estimated, and found to be small.

\end{abstract}

\pagebreak

\section*{\underline{Introduction}}
	  
	Recent measurements of the flavor \cite{Ama91,NA51,E866} and 
spin \cite{2} dependence 
of quark distributions in the nucleon have led to a revival of interest in
the soft QCD physics that determines the shape and normalization of parton
distributions \cite{3}, and has led to a re-examination of some of the
fundamental assumptions embodied in the parametrizations \cite{4} used to 
describe data from high energy experiments. In particular, the violation
of charge symmetry in the valence quark distributions has been calculated 
by a 
number of authors \cite{5,6,6a,6b} in the context of quark 
models, and surprisingly large violations, as large as 5-10\%
, have been found at large $x_{Bj}$ \cite{5}. Although charge 
symmetry is assumed in all phenomenological parton distributions, 
there is not a great deal of direct experimental evidence which justifies
this assumption. The strongest upper limit on parton charge symmetry
violation(CSV) can be obtained by comparing the $F_2$ structure
function for charged lepton deep inelastic scattering with the $F_2$
structure function measured in neutrino-induced charged current
reactions.  The CCFR group has recently carried out such a test 
using the most recent available data \cite{Sel97}.  They compared
their neutrino cross sections on iron with the $F_2$ structure
functions extracted from muon-deuterium measurements of the
NMC group \cite{Ama91}.  In the region $ 0.1 \le x \le 0.3$, these
recent experiments can place upper limits of about 6\% on parton
CSV.  For larger $x$ the upper limits are substantially larger, 
while for $x < 0.1$ the present data appears to indicate a 
substantial violation of charge symmetry.  A number of experiments
have been suggested which would measure directly charge symmetry
violation in parton distributions \cite{Lon97}. 
       			
	In this paper, we combine the approaches of Refs.\ \cite{5} and 
\cite{6} and
examine the violation of charge symmetry in the valence and sea quark 
distributions of the nucleon in a manner independent of the choice of 
any particular quark model. In the following section, we present the 
formalism 
used to produce parton distributions from quark model wavefunctions, and
examine the variation of those distributions with small changes in the
mass of the system of spectator quarks. In the next section, we use
these results to examine the changes in the valence quark distribution
due to SU(4) violation via color magnetism, and charge symmetry violation 
via quark masses and electro-magnetism. Combining these results, we obtain
a wavefunction independent prediction relating the dominant contribution to
 the CSV valence distribution 
to the well known difference between the $u$ and $d$ valence 
distributions.
In the penultimate section, we estimate the charge symmetry violating sea 
quark 
distributions, assuming that the dominant CSV effect comes from the 
mass and electro-magnetic differences of the 
participating quarks, rather than from dependence of the quark
wavefunctions on charge symmetry, as is the case for the valence
quark CSV contributions \cite{5}. Additionally, we make simple estimates of 
the contribution of 
CSV to the Gottfried Sum Rule. Finally, we compare our results with those 
of other authors and discuss their implications.
  
\section*{\underline{Quark Distributions from Quark Models}}

	For the purposes of this paper, we shall adopt the Adelaide
method for calculating quark distributions from model wavefunctions 
\cite{5}.
The starting point for the method is the reduction of quark distributions
to the form \cite{9} 
\begin{equation}
q(x)= M \int {d^3k\over (2\pi)^3} \vert\Psi_+(k)\vert^2 \delta(E_q-k_3-Mx)
\label{one}
\end{equation}
where $x$ is the Bjorken scaling variable, $M$ is the mass of the target,
$\Psi_+(k)$ is the light cone wavefunction for the struck quark in
momentum space, and $E_q$ is the energy of the struck quark. An analogous 
expression gives the antiquark distributions 
\begin{equation}
\overline q(x)= M \int {d^3k\over (2\pi)^3} \vert\Psi^\dagger_+(k)\vert^2 
\delta(E_q-k_3-Mx).
\label{two}
\end{equation}

The essence 
of the Adelaide approach lies in the kinematic assumption that the three
quark system can be divided into the struck quark plus a diquark spectator 
system, and that the distribution of masses of the spectator system is
sufficiently sharply peaked that the diquark can be thought of as an
on-shell system with a definite mass. This completely specifies the 
kinematics of the problem and allows the quark energy appearing in the 
delta functions of Eqs.\ \ref{one} and \ref{two} to be replaced by 
the difference
between the target mass and the on-shell diquark energy. Since the direct 
calculation of these distributions is not our interest here, we will not 
reduce this expression further, but simply note that the expression for
the quark distribution can be reduced to a single integral over the 
momentum space wavefunction of the struck quark \cite{5}, and is guaranteed
to have the correct support as a function of $x_{Bj}$ regardless of the
properties of the wavefunctions used to evaluate the integral.  

\subsection*{\underline{Symmetry Breaking}}   

With the exception of Ref. \cite{6}, the approach to study the 
violation of 
parton distribution symmetries has been to calculate entire distributions
in the framework of a given quark model and then to compare the results
from small variations in the model parameters. In Refs. \cite{5} and 
\cite{10}, the dependence of the valence quark distribution on the model 
wavefunction and diquark mass parameter has been studied in the context of 
examining charge and flavor symmetry in the valence sector. The dependence 
of the valence distributions on the details of the quark wavefunction was 
found to be small in comparison to the effects generated by changes in the 
diquark mass parameter. Similar conclusions were reached in Ref.\ 
\cite{6a}, where charge symmetry 
breaking effects were studied using a different approach to connect quark 
model wavefunctions to valence distributions. Based on these results, we 
shall assume that the dominant contribution to the breaking of parton 
distribution symmetries is generated not by the changes in struck quark 
wavefunctions, but by the changes in the kinematic constraints brought
about by shifts in the masses of the spectator quark systems and the 
nucleon
itself. This allows a quark model independent prediction for the change in 
parton distributions when the symmetry is broken. To see this, begin with 
Eq.\ \ref{one} for the valence 
quark distribution, and make a small variation in the mass of the spectator 
diquark, $M_d\rightarrow M^0_d+\delta M_d$, where $M^0_d$ is the 
diquark mass
in the symmetric limit, and $\delta M_d$ is the shift in the diquark mass
due to symmetry breaking.
Expanding Eq.\ \ref{one} to first order in $\delta M$, we obtain
\begin{equation}
\delta q(x) = - M \int d^3k \vert\Psi_+(k)\vert^2 \frac{\delta M_d M^0_d}
{E^0_d}\delta^\prime(M(1-x)+k_3-E^0_d({\bf k})),
\label{three}
\end{equation}    
where quantities with the superscript $0$ are taken in the symmetric limit,
and $E_d({\bf k})$ is the diquark energy. Changing the derivative of the 
delta function to a derivative in $x$, using kinematics to solve for 
$E^0_d$,
and neglecting the transverse momentum of the diquark\cite{10a}, one finally 
obtains
\begin{equation}
\delta q(x) \approx \delta M_d \frac{d}{dx}
\bigg[ \frac{2M^0_d(1-x)}{M^2(1-x)^2+ (M^0_d)^2}\, q^0(x) \bigg].
\label{dqanal}
\end{equation}
As we shall demonstrate in the following, this expression has two
important advantages over explicit model realizations: First, as
in Ref.\ \cite{6},  
$\delta q(x)$ is determined in terms of $q^0(x)$,so the change in the
distribution can be extracted from measured distributions rather than 
models. Second, the size of the change is controlled by the diquark mass
shift, which allows us to relate different symmetry breaking effects 
to one another.

\section*{\underline{Symmetry Breaking in the Valence Sector}}

	In models where confinement is implemented via interactions which 
are independent of both spin and flavor, the wavefunctions of the up and 
down valence quarks are identical, leading via Eq.\ \ref{one}  
to the prediction
\begin{equation}
\frac{d_v(x)}{u_v(x)}=\frac{1}{2},
\label{five}
\end{equation}
where $u_v(x)$ and $d_v(x)$ are respectively the up and down valence 
quark distributions, and the factor of two merely expresses the relative 
normalization of the 
distributions. These relations, which relate distributions with different
flavors within the same hadron, are somewhat inappropriately termed 
flavor symmetries \cite{11}. Such symmetries are a consequence of 
dynamical assumptions about the nature of confinement in QCD. In this
instance, the quark model SU(4) spin-isospin symmetry gets broken
by the color hyperfine interaction, leading to the well-known
dominance of $u_v$ over $d_v$ at large $x$. 
	
	This symmetry breaking has been modeled by Close and Thomas 
\cite{10}, who considered the spin-flavor correlations
in the SU(4) nucleon wavefunction and the mass splitting of spin 
1 and spin 0 diquark pairs brought about by the color hyperfine
interaction. In this picture, the valence distribution depends 
not on the flavor of the struck quark, but rather on the spin
of the spectator diquark system. Explicitly,
\begin{eqnarray}
u_v(x)&=&\frac{3}{2} q_v^{S=0}(x) + \frac{1}{2} q_v^{S=1}(x),\nonumber\\
d_v(x)&=& q_v^{S=1}(x), 
\end{eqnarray}
where the superscript refers to the spin of the diquark spectator. In
the SU(4) symmetric limit, $q_v^{S=0}(x)=q_v^{S=1}(x)$ and Eq.\ \ref{five} 
is recovered. Color hyperfine effects are included via a Hamiltonian
of the form 
\begin{equation}
H_{hf}=v\sum_{i>j}\sum_{a=1..8} {\bf \sigma_i\cdot\sigma_j}\lambda^a_i
\lambda_{aj},
\end{equation}
with $\frac{1}{2}{\bf \sigma_i}$ the spin of quark $i$, $\lambda^a_i$ the
corresponding color generator, and $v=75$ MeV is normalized by the 
$N-\Delta$ splitting. The diquark masses are shifted according to
\begin{equation}
\delta_{hf} M_d^{S=1}=-\frac{1}{3}\delta_{hf} M_d^{S=0}=\frac{2}{3}v,
\end{equation}  
so that, from Eq.\ \ref{dqanal}, 
\begin{equation}
\delta_{hf} q_v^{S=1}(x)=-\frac{1}{3}\delta_{hf} q_v^{S=0}(x)
\end{equation}
and 
\begin{equation}
\delta_{hf} q_v^{S=1}(x)=\frac{2d_v(x)-u_v(x)}{6}.
\end{equation}
Interestingly, this pattern of symmetry breaking predicts that the valence
distributions measured in lepto-production from neutron targets are 
precisely the SU(4) symmetric distributions.

	Now we turn our attention to the case of charge symmetry 
violation, which relates distributions of opposite isospin in
targets of opposite isospin. In the charge symmetric limit,
\begin{eqnarray}
u_v^p(x)&=&d_v^n(x)\nonumber\\
d_v^p(x)&=&u_v^n(x),
\end{eqnarray}
where for example $u_v^p(x)$ denotes the distribution of $u$ quarks in the 
proton, $d_v^n(x)$ is the distribution of $d$ quarks in the neutron, and 
we have analogous definitions for the minority quark distributions.
	
	At the quark level, charge symmetry is violated by quark mass 
and electro-magnetic effects and is generally expected at the level
of one per cent. Since both of these effects are iso-vector, there
is no CSV contribution to the mass of the isoscalar diquark, and we may
write, for the majority quark distributions, 
\begin{eqnarray}
d_v^p(x)&=& q_v^{S=1}(x)+\frac{1}{2}\delta_\CSV q_v^{S=1}(x) \nonumber\\
u_v^n(x)&=& q_v^{S=1}(x)-\frac{1}{2}\delta_\CSV q_v^{S=1}(x)  ,
\label{twelve}
\end{eqnarray} 
where $\delta_\CSV q_v^{S=1}(x)$ is the change in the quark distribution
generated by the shift in the $I_3=\pm 1$ diquark masses via Eq.\ 
\ref{dqanal}. Since
the only difference between the CSV case and the flavor/SU(4) symmetry 
violating distribution is the value of $\delta M_d$, the two distributions 
are the same up to a normalization, so that
\begin{equation}
\delta_\CSV q_v^{S=1}(x) =\frac{\delta_\CSV M_d}{\delta_{hf}M_d} 
\delta_{hf}
q_v^{S=1}(x)= \frac{\delta_\CSV M_d}{\delta_{hf}M_d}
\left( \frac{2d_v(x)-u_v(x)}{6}\right)  .
\label{analyt}
\end{equation}

From the bag model, a robust estimate of the isovector diquark mass 
splitting
is -4 MeV \cite{12}, yielding $\delta_\CSV M_d/ \delta_{hf}M_d \approx
.08$. Using the CTEQ4 LQ  distributions \cite{4} to parametrize the 
valence 
distributions, the charge symmetry violating contribution to the minority
quark valence distribution due to the shift in the diquark mass is shown in 
Fig. 1. We plot the quantity $x\delta d_v(x) = x\left( d^p_v(x) - u^n_v(x)
\right)$, where the CSV terms are calculated from Eq.\ \ref{analyt}. 
Also shown are result of a bag model calculations using two
different approaches to extract the quark distributions from model
wavefunctions \cite{6b}.  The reasonable agreement between the analytic
result of Eq.\ \ref{analyt} and the two different model calculations
of CSV gives us confidence that the model calculations give rather
robust predictions of valence quark charge symmetry violation.

\begin{figure}
\centering{\ \psfig{figure=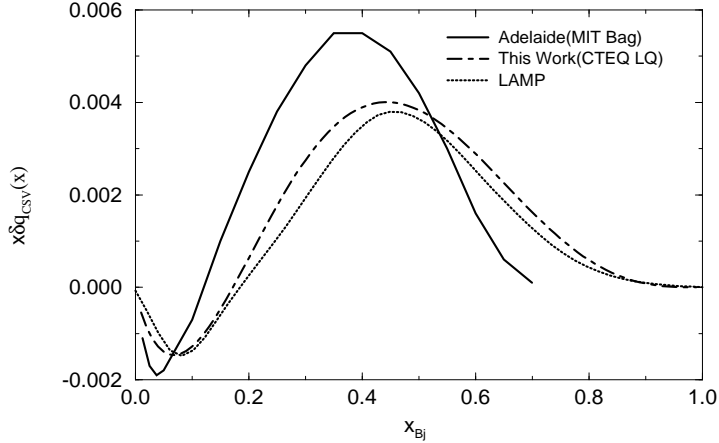,width=10cm}}
\vspace{.1in}
\caption{Diquark mass contribution to Charge Symmetry violating quark 
distributions for minority valence quarks.  Curves show 
$x\left(d_v^p(x)- u_v^n(x)\right)$ defined from Eqs.\ \protect\ref{twelve}
and \protect\ref{analyt}. 
Curves were calculated using the CTEQ4 LQ distribution at $Q^2=0.49$ 
GeV$^2$.
Also shown are model calculations from Refs.\ \protect\cite{5} and 
\protect\cite{6a}.}
\end{figure}

	To complete the calculation of the CSV distributions we must 
include the effect of the proton-neutron mass difference. This proceeds
in essentially the same manner outlined in the last section, with the
slight subtlety that the variation with respect to the target mass
should be taken keeping the product $Mx$ fixed, as the mass is implicit
in the definition of $x$ (see Ref.\ \cite{6a} for an explanation of this 
point).  The resulting addition to the CSV distribution yields
\begin{eqnarray}
\delta d_v(x) &\equiv& d_v^p(x)-u_v^n(x) = 
  .013(u_v(x)-2d_v(x))-\frac{\delta M}{M}\frac{
 d\,d_v(x)}{d x}\nonumber\\  \delta u_v(x) &\equiv&   	  
u_v^p(x)-d_v^n(x) = -\frac{\delta M}{M}\frac{d\, u_v(x)}{d x},
\label{csvdef}
\end{eqnarray}
where $\delta M=-1.3$ MeV is the proton-neutron mass difference.The 
resulting
CSV distributions (multiplied by $x$) are shown in Fig. 2 for the 
CTEQ4 LQ 
distributions, along with the CSV contribution to the minority quark 
distribution ($x\delta d_v(x)$) from Fig. 1 (the dash-dot curve in
Fig.\ 2).

\begin{figure}
\centering{\ \psfig{figure=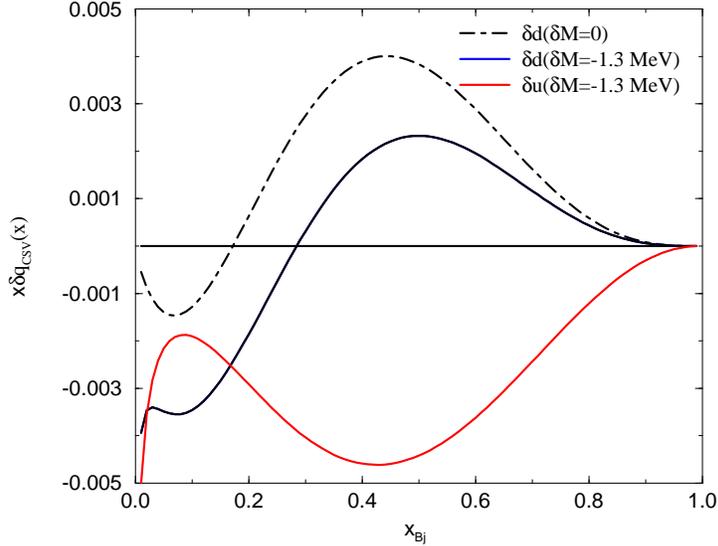,width=10cm}}
\vspace{.1in}
\caption{Charge Symmetry Violating quark distributions for majority
($\delta u$) and minority ($\delta d$) valence quarks.  Solid curves 
include effects of both $n-p$ mass difference and diquark mass 
contribution; 
dot-dashed curve shows minority quark CSV neglecting the $n-p$ mass 
difference.  Curves were 
calculated using the CTEQ LQ distribution at $Q^2=0.49$ GeV$^2$.}
\end{figure}

As a consequence of the singularities in the CTEQ parametrizations, the 
nucleon mass correction diverges non-integrably at $x=0$, cf.\ Eq.\ 
\ref{dqanal}. This reflects the 
inadequacy of the diquark spectator approximation in this region, which is
dominated by states having one or more pairs of sea quarks. At larger $x$,
where the calculation is more reliable,the nucleon mass correction 
partially 
cancels out the change due to the diquark mass shift, a result which can be 
expected on the basis of Eq.\ \ref{three}. In accordance with the results of 
Refs.\ \cite{5,6,6a}, the
magnitude of the CSV contribution to the minority and majority quark 
distributions are roughly comparable, so that the smaller minority quark 
distribution is more sensitive to charge symmetry violation than the majority 
distribution.
  
\section*{\underline{Charge Symmetry Violation in the Nucleon Sea}}

There has been much interest in the question of flavor symmetry violation 
in the nucleon sea, following the precise measurement of the 
Gottfried sum rule \cite{Got67} by the NMC group \cite{Ama91}.  The 
Gottfried sum rule is obtained by integrating the $F_2$ structure
functions for lepton-induced deep inelastic scattering(DIS) on protons and 
neutrons, 
\begin{eqnarray} 
S_G &=& \int_0^1 {dx\over x} \left[ F_2^{\mu p}(x) - F_2^{\mu n}(x)
  \right] \nonumber \\ &=& {1\over 3} + {8\over 9} \left( P_{\bar{u}/p} 
  - P_{\bar{u}/n} \right) + {2\over 9} \left( P_{\bar{d}/p}  - 
  P_{\bar{d}/n} \right) ~~, 
\label{gottf}
\end{eqnarray} 
where we define the sea quark probabilities as 
\begin{eqnarray*}
 P_{\bar{u}/p} &\equiv& \int_0^1 \bar{u}^p (x)\, dx ~~.
\end{eqnarray*}
In deriving Eq.\ \ref{gottf} we have used the normalization of
the valence quark distributions; if we assume parton charge symmetry 
we obtain the ``standard'' form for the Gottfried sum rule, 
\begin{eqnarray*}
 S_G &=& {1\over 3} - {2\over 3}\left( P_{\bar{d}/p}  - P_{\bar{u}/p} 
  \right) 
\end{eqnarray*}

By measurements over the range $0.004 \le x \le 0.8$ for muon DIS on 
proton and deuteron targets, and extrapolation over the unmeasured 
region, the NMC group obtained $S_G = 0.235 \pm 0.026$.  If one 
assumes charge symmetry, this result implies an asymmetry in the 
probabilities for finding up and down sea quarks in the proton, 
\begin{eqnarray*}
 P_{\bar{d}/p}  - P_{\bar{u}/p} &=& 0.147 \pm 0.039 ~~. 
\end{eqnarray*}
However, as was pointed out by Ma \cite{13}, we could alternatively 
assume flavor symmetry but not charge symmetry for the sea quark 
distributions, in which case the NMC result would imply 
\begin{eqnarray*}
 P_{\bar{u}/p}  - P_{\bar{u}/n} &=& -0.088 \pm 0.023 ~~. 
\end{eqnarray*}
Experimental upper limits on sea quark charge symmetry cannot
at present rule out parton CSV contributions.  So it is possible that
the Gottfried sum rule result arises either from CSV effects, or from 
a combination of flavor symmetry and charge symmetry violation.   

In this section we will estimate the magnitude of charge symmetry
violation for sea quarks, and the contribution this might make
to the Gottfried sum rule.  The effect of flavor symmetry or charge 
symmetry violation on the sea quark distributions
is complicated by the fact that there are no sum rules constraining the 
normalization of the sea quark distributions. Indeed, in the standard
parametrizations \cite{4} at high momentum scales the total number of sea 
quarks is infinite, a fact that has been used to suggest that 
fractionally small CSV in the sea may lead to sizeable contributions to 
the Gottfried sum 
rule \cite{13}. Since this contribution, regardless of its source, is
independent of the momentum scale at which it is measured \cite{14}, it is 
amenable to modeling at a relatively low scale, say $Q^2 = .5$ GeV$^2$, 
where it has been argued that the number of sea quarks in the nucleon is 
finite \cite{15}.

\subsection*{\underline{Sum Rule Contributions}}

The CSV contributions to the sea separate, somewhat artificially,
into two classes: Strong CSV effects which change the number of sea quarks 
in the nucleon (and potentially contribute to the Gottfried sum rule), and 
weak CSV effects which  
change the shape of the sea distributions without altering their
normalization. We shall begin by considering the first of these classes,
since the extent to which the Gottfried sum rule violation is a reflection 
of CSV rather than flavor symmetry violation may provide an important 
constraint on the second class of CSV contribution. The simplest 
perturbative 
estimate of the size of the CSV contribution is obtained by modeling 
the nucleon as a valence state coupled to a small number of states with an 
additional quark-antiquark pair
\begin{equation}
\vert N> = Z (\vert N_{val}> +\sum_\alpha A_{\alpha/N} 
\vert N_{val}q_\alpha\bar q_\alpha>),    
\end{equation}
with $Z$ a normalization, $A_\alpha$ the amplitude for finding the 
``extra''
quark anti-quark pair of flavor $\alpha= u,d,s$, and $N_{val}$ is a 
three
quark state with the same third component of isospin as the original 
three 
quark state. Assuming flavor symmetry, the amplitude for creating an 
extra 
quark pair of flavor $\alpha$ is given, in perturbation theory, by
\begin{equation}
     A_{\alpha/N} \approx {\lambda\over (M_{5Q\alpha}-M_{nuc})},
\end{equation}
where $M_{5Q}$ is the mass of the 4 quark, anti-quark state, and $\lambda$
is a typical hadronic mass scale determined by the strong coupling 
constant
 and the details of the wavefunction.If charge symmetry is also good, 
 this 
amplitude is independent of the flavor of the quark-antiquark pair, so 
that 
the probability for creating a quark anti-quark pair of flavor $\alpha$ 
is  
\begin{equation}
P_{\alpha/N}\approx {\lambda^2\over(M_{5Q\alpha}-M_{nuc})^2}, 
\end{equation}
which, apart from Pauli effects which violate the flavor symmetry we've 
assumed, is essentially the result of Donoghue and Golowich \cite{16}.

	To incorporate charge symmetry violation into this picture
requires only an estimate of the mass shift due to quark mass and 
electro-magnetic effects. We parametrize this difference via a term that
counts the number of up and down quark masses in the composite system,
and through the inclusion of an electro-magnetic contribution to the 
energy
which we assume is proportional to the sum of the pairwise products of the 
charges of the constituents. The quark mass difference and electro-magnetic 
energy are then fit to the mass shifts of the diquark and nucleon, and the 
results are extrapolated to systems with more quarks. This procedure 
clearly neglects a great deal of the physics of these systems, but should 
be sufficient to provide an estimate of the energy scales involved.The 
resulting mass shifts between charge conjugate systems can then be written
\begin{equation}
\delta M_\CSB= (N_u- N_d)(m_u-m_d)-\delta\Sigma\, \epsilon_\EM,
\end{equation}
where $N_f$ is the number of quarks plus antiquarks of flavor $f$ in the
system in question, $\delta\Sigma$ is the change in the sum of the 
products 
of charges under charge conjugation, and $\epsilon_\EM$ is the average
electro-magnetic energy of each quark pair. The mass shifts that result 
from 
this procedure are shown in Table 1.

Expanding to first order in the symmetry violating terms yields 
relations between the number of sea quarks of charge conjugate flavors 
in the proton and neutron:
\begin{eqnarray}
P_{\bar u/p}&=&P_{\bar d/n}\left 
(1-{2(m_u-m_d)-\frac{1}{3}\epsilon_\EM\over m_q}\right )
\nonumber\\
P_{\bar d/p}&=&P_{\bar u/n}\left 
(1+{2(m_u-m_d)-\frac{1}{3}\epsilon_\EM\over m_q}\right )
\nonumber\\
P_{\bar s/p}&=&P_{\bar s/n},
\label{conjg}
\end{eqnarray}
where $m_q=360$ MeV is the constituent quark 
mass in the symmetric limit.
Remarkably, since the energy required to make a pair of a particular flavor
is the same for both the proton and neutron in this parametrization, there 
is no contribution to the Gottfried Sum Rule in the flavor symmetric 
limit. In a more sophisticated treatment of quark mass and 
electro-magnetic effects we expect that the CSV
contribution will be non-zero, but since a contribution to the Gottfried 
Sum Rule needs top be odd under the combined operation of flavor and charge
symmetry, which requires additional CSV and Flavor violating effects, the CSV 
contribution to the Gottfried sum rule will be suppressed below the natural
scales for CSV effects arising from quark
mass differences and electromagnetic terms (proportional to 
$(m_u-m_d)/m_q$ and $\alpha$, respectively).

	Another source for CSV in the sea is the possibility of charge 
dependence of the quark gluon coupling constant. By dressing the 
quark-gluon
vertex with a photon as in Figs. 3a and 3b, the effective quark-gluon 
coupling
for $u$ quarks is different than that for $d$ quarks \cite{18}, so that 
the
probability of creating a $q \bar{q}$ pair becomes dependent on both the 
flavor
of the pair created and the flavor of the quark that emitted the gluon.

\begin{figure}
\centering{\ \psfig{figure=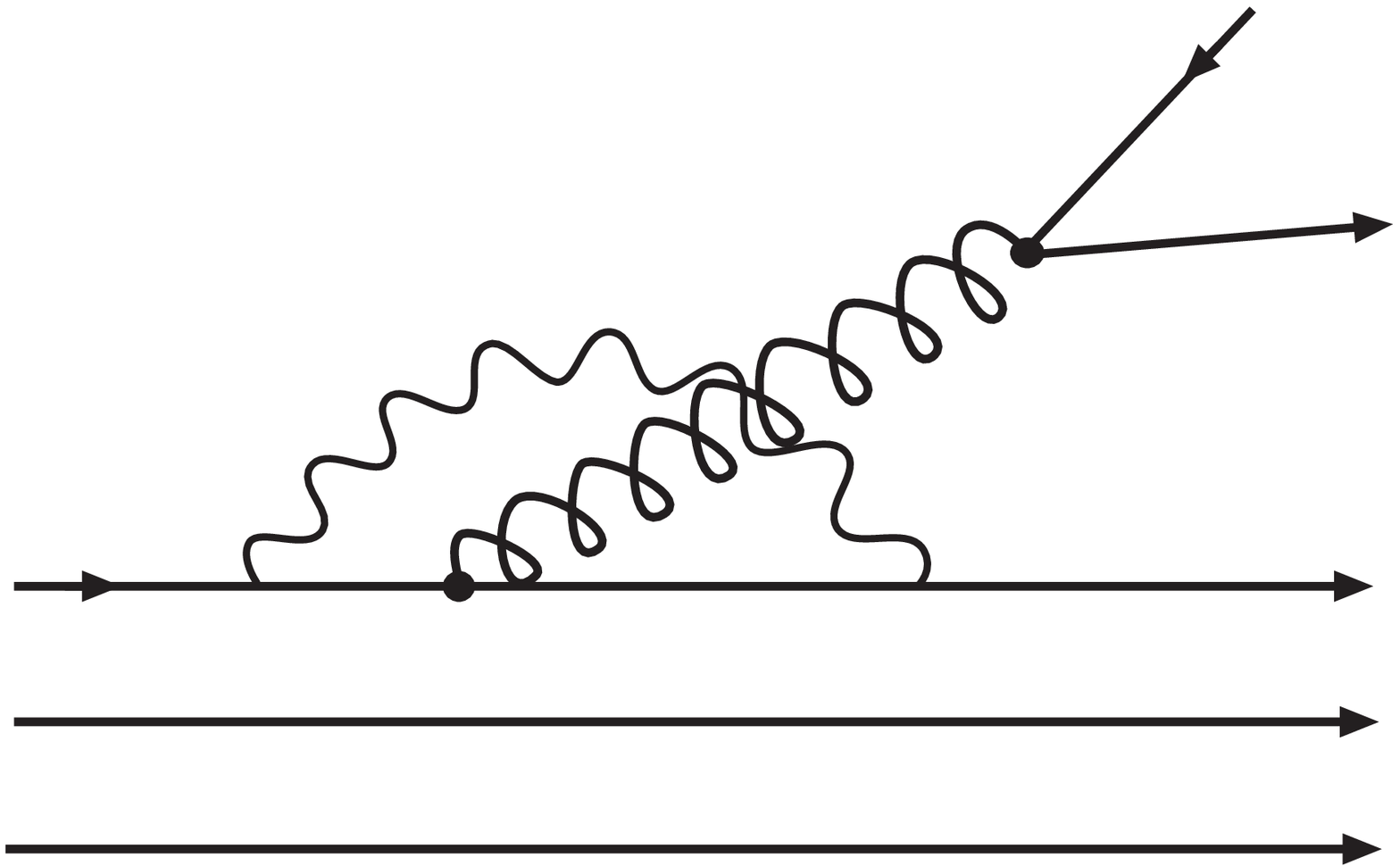,width=5cm}\hspace{2cm}
\psfig{figure=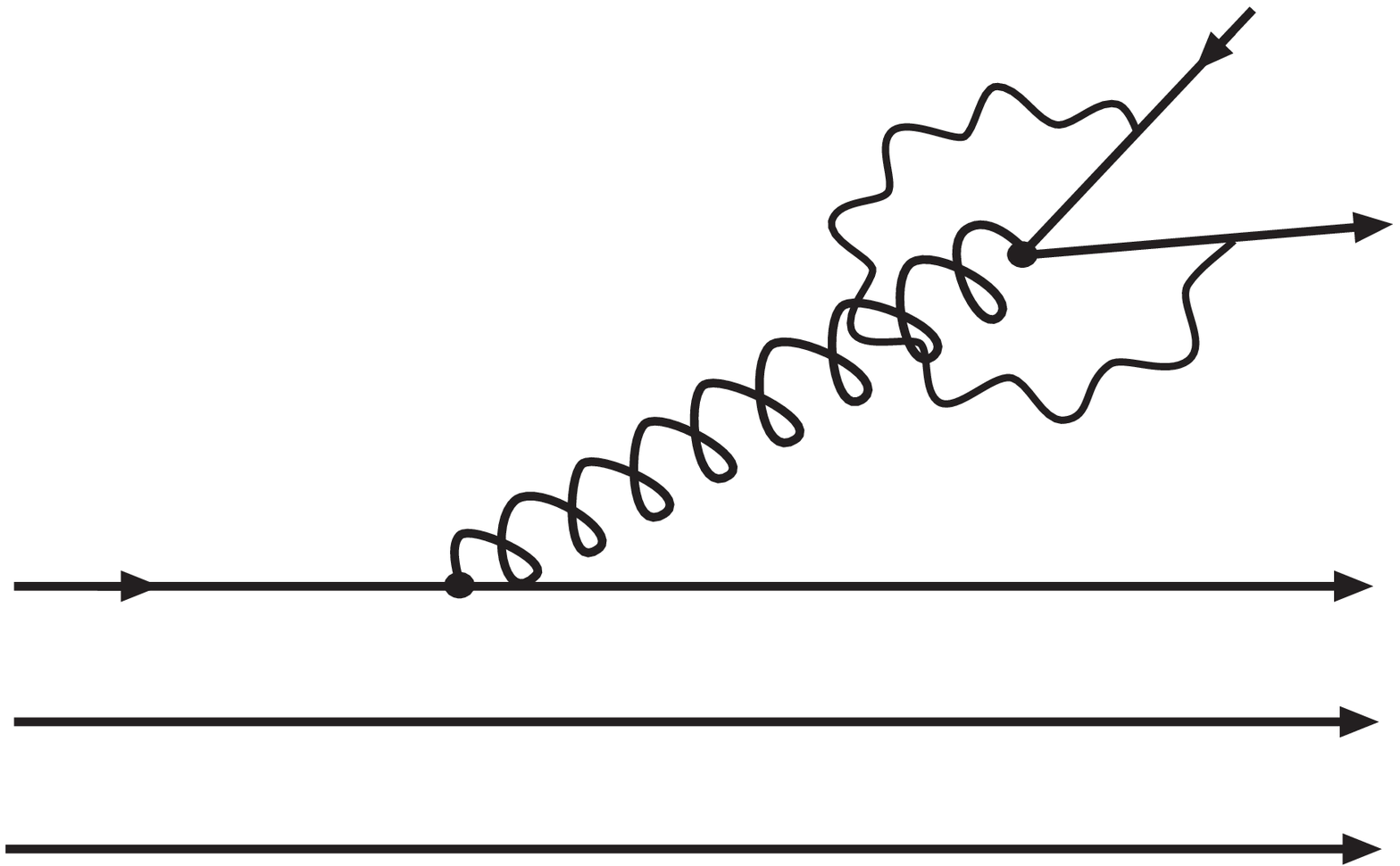,width=5cm }}
\vspace{.1in}
\caption{ Photons dressing the quark-gluon vertex during the production
 of sea quarks. The probability of producing a sea quark pair will depend 
 on 
the charge of the created quarks(Fig. 3b) and on the quark that initially 
produced the gluon(Fig. 3a).}
\end{figure}
 
At low momentum scales, the effective gluon coupling to a quark of flavor 
$f$ is given by
\begin{equation}
\alpha_{sf}^{eff}=\alpha_s(Q^2)(1+e_f^2V(b)),
\end{equation}
where $\alpha_s$ is the strong coupling constant, $b=\Lambda^2/Q^2$,$e_f$ 
is the quark charge,  and
\begin{equation}
V(b)=\frac{\alpha}{4\pi}(-1-\frac{\pi^2}{3}-\ln{b}-\ln^2b),
\end{equation}
with $\alpha$ the electro-magnetic coupling. Rewriting this, we get
\begin{eqnarray}
\alpha_s^u&=& \alpha_{s0}+\delta\alpha_s\nonumber\\
\alpha_s^d&=& \alpha_{s0}-\delta\alpha_s,
\end{eqnarray}	
where now 
\begin{equation}
\delta\alpha_s={\alpha_s(Q^2)\over 6}V(b)
\end{equation}
is the charge asymmetric piece of the strong coupling and
$\alpha_{s0}=\frac{1}{2}(\alpha_s^u+\alpha_s^d)$. 

	Assuming that the sea quarks at low momentum scales are
generated by the valence quarks, we have
\begin{eqnarray}
P_{\bar u_p}&\propto & (2\alpha_s^u+\alpha_s^d)\alpha_s^u\nonumber\\
P_{\bar u_n}&\propto & (\alpha_s^u+2\alpha_s^d)\alpha_s^u\nonumber\\
P_{\bar d_p}&\propto & (2\alpha_s^u+\alpha_s^d)\alpha_s^d\nonumber\\
P_{\bar d_n}&\propto & (\alpha_s^u+2\alpha_s^d)\alpha_s^d,
\end{eqnarray} 
from which we obtain, to first order in the CSV terms,
\begin{eqnarray}
P_{\bar u/n}&=& P_{\bar d/p}\nonumber\\
P_{\bar d/n}&=& P_{\bar u/p}(1-4{\delta\alpha_s\over\alpha_{s0}}).
\end{eqnarray} 
	
	Since the valence quarks provide a source of flavor asymmetry,
there is a contribution to the Gottfried Sum Rule. Using the CTEQ 
distributions to calculate $P_{u/p}=.365$ and $\Lambda=200$ MeV, we find 
the CSV contribution to the Gottfried sum rule to be $6.3\times 10^{-4}$, 
which is far too small to
explain the observed deviation. Similarly, there is a small excess of 
strange 
quarks in the proton over those in the neutron, according to
\begin{equation}
P_{\bar s/p}-P_{\bar s/n} = 
+2{\delta\alpha_s\over\alpha_{s0}}P_{\bar s/p} \approx 2\times 10^{-4}.
\end{equation}

Having eliminated the possibility of a large CSV contribution to
the Gottfried sum rule, we return briefly to the possibility that the 
largest part of the sum rule comes from violation of flavor symmetry, and
that the violation of flavor symmetry leads to a CSV contribution to the 
sum as well. Combining Eq.\ \ref{conjg} with a flavor asymmetric sea yields
\begin{equation}
\delta_\CSV S_G= \frac{5}{9} ({2(m_u-m_d)-
\frac{1}{3}\epsilon_\EM\over m_q})
(P_{\bar d/p}-P_{\bar u/p})\approx -.0012,
\end{equation}
using the CTEQ parton distributions.  Our estimated CSV contribution is 
still far too small to affect significantly 
the extraction of the flavor asymmetry. We conclude that the 
prospects for a large contribution to the Gottfried sum rule from sea 
quark CSV are extremely slim, at least within the assumptions in
our calculation of charge symmetry violating effects. 

\subsection*{\underline{Weak Charge Symmetry Violation in the Sea}}
		  
Since we have been unable to produce a mechanism for generating a
CSV distribution whose integral is significantly different from zero,
we are led to investigate the possibility of weak charge symmetry 
breaking,
where the shapes of the sea distributions in the neutron and proton are 
different, but their normalizations remain the same. By analogy with 
the CSV effects in the valence distributions, we shall proceed by 
postulating that the 4 spectator partons can be assumed to belong to
a tetra-quark, and that the mass of the recoiling tetra-quark can be 
assumed 
to be approximately constant with a value roughly equal to the sum of the 
constituent quark masses. Since there is a greater variety of states 
available
to the four parton spectator system, this assumption is much less certain 
than the analogous assumption in the valence case, but it will allow for
an exploration of the magnitudes involved in the problem. 

The basic argument proceeds as in the case of the valence 
distributions, by calculating the shift in the tetra-quark mass produced
by the symmetry violating interactions and expanding to get   

\begin{equation}
\delta\bar q(x)=\bar q^p(x)-\bar q^n(x) = \delta M_T \frac{d}{dx}
\bigg[ \frac{2M^0_T(1-x)}{M^2(1-x)^2+ (M^0_T)^2}\,\bar q^0(x) \bigg],
\end{equation}
where $M^0_T$ is the charge symmetric tetra-quark mass, and $\bar q_0(x)$
is the anti-quark distribution in the symmetric limit. For the light quark
sea, we have used a tetra-quark mass of $1440$ MeV, roughly four times the
constituent quark mass. If, as many models assert, pionic effects dominate
the sea at these low scales, this assumption will underestimate the size 
of 
the CSV contribution to the light quark sea. For the systems containing 
strange quarks, we have assumed a constituent quark mass of 500 MeV.  
 In addition to this correction, there
will be a correction due the change in the nucleon mass introduced by CSV,
which is calculated in the same fashion as the analogous contribution for 
the
valence quarks.The values of the CSV induced mass shifts for strange and 
nonstrange tetra-quarks are listed in Table 1, and the resulting CSV sea 
distributions are shown in Fig. 4. 

\begin{figure}
\centering{\ \psfig{figure=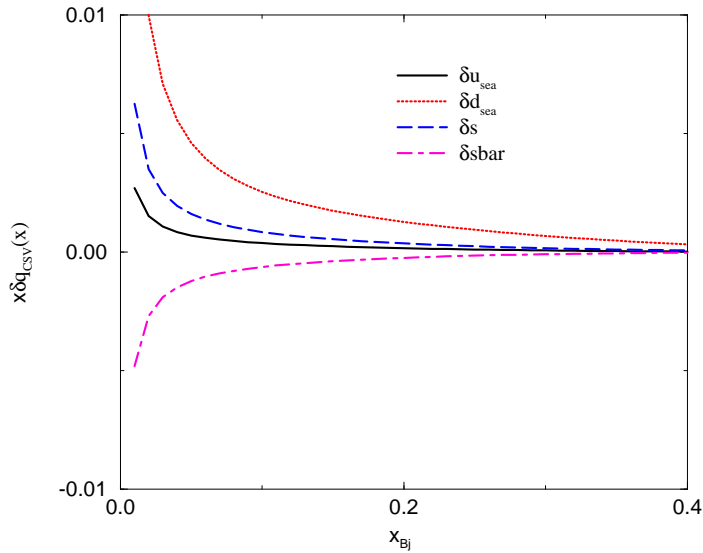,width=10cm}}
\vspace{.1in}
\caption{Charge symmetry violating sea quark distributions. The solid, 
dotted,
dashed, and dot-dashed curves are 
$x\delta\bar{u} =x(\bar u^p-\bar d^n)$, $x\delta \bar{d} =x(\bar d ^p-
\bar u^n)$, $x\delta s = x(s^p-s^n)$, and $x\delta\bar s=x(\bar s^p-
\bar s^n)$, respectively.}
\end{figure}

Just as in the case of the valence distributions, the large contribution
to sea quark asymmetries at very small $x$ is a consequence of the 
singular form of the parton distributions in this region, and indicates
the breakdown of the tetra-quark picture we've used here.  At larger $x$,
the relative size of $\delta\bar{d}$  over $\delta\bar{u}$ reflects the 
fact that
the tetra-quark spectator has $I_3=0$ when a $\bar d$ is struck, so that 
there is no tetra-quark mass shift to partially cancel the nucleon mass 
shift,
as is the case when $\bar u$ is struck. For the strange CSV distributions,
a similar cancellation occurs for the spectator tetra-quark when an $s$ 
quark
is struck, but since this cancellation depends sensitively on the 
parameters
chosen in Table 1, we conclude this result may be highly model dependent 
and
therefore untrustworthy. In general, we do not find surprisingly large 
contributions for any of the sea quark distributions.

\section*{Summary} 

In this paper, we have used the Adelaide approach to calculating 
quark model parton distributions to calculate the effect of CSV on the 
parton distributions of the proton. For the valence quarks, we verify in 
a 
model independent fashion the anomalously large CSV effects observed in 
the 
minority valence quark distribution by several authors \cite{5,6,6a}, and 
we relate these distributions to the difference between the $u$ 
and $d$ valence distributions.  Since there is little data to constrain
the size of the CSV distributions, proposed tests of CSV \cite{Lon97} will
provide a sensitive test of the physics contained in QCD-inspired 
models of the nucleon.  

Additionally, we have made estimates of the size of CSV effects in
the nucleon sea.  First, we assume a strong violation of charge
symmetry, which would alter the value of sum rules which assume
charge symmetry.  For the Gottfried sum rule, there is no CSV 
contributions unless there is a simultaneous violation of flavor
symmetry in either the sea or valence quarks.  In either case, we
estimate that the CSV contributions to the Gottfried sum rule are 
much smaller than are suggested by experiment.  We have also 
estimated the weak CSV contribution to the nucleon sea, using
methods similar to the valence quark CSV calculations.  Again, we
find the charge symmetry violating distributions are extremely
small, and highly unlikely to make significant contributions to
any observable.    

\subsection*{Acknowledgments}

The authors wish to acknowledge support from NSF Research 
Contract No. PHY-9722706, and to D. Murdock for assistance in producing 
the figures, and A.W. Thomas for useful discussions of alternate methods 
for arriving at an analytic expression for the CSV distributions.

\pagebreak
\begin{centering}
\centerline{\underline{Figure Captions}}
\end{centering}
\begin{itemize}
\item{} Fig. 1 -  Diquark mass contribution to Charge Symmetry violating 
quark 
distributions for minority valence quarks.  Curves show 
$x\left(d_v^p(x)- u_v^n(x)\right)$ defined from Eqs.\ \protect\ref{twelve}
and \protect\ref{analyt}. 
Curves were calculated using the CTEQ4 LQ distribution at $Q^2=0.49$ 
GeV$^2$.
Also shown are model calculations from Refs.\ \protect\cite{5} and 
\protect\cite{6a}.
\item{} Fig. 2 - Charge Symmetry Violating quark distributions for majority
($\delta u$) and minority ($\delta d$) valence quarks.  Solid curves 
include effects of both $n-p$ mass difference and diquark mass contribution; 
dot-dashed curve shows minority quark CSV neglecting the $n-p$ mass 
difference.  Curves were 
calculated using the CTEQ LQ distribution at $Q^2=0.49$ GeV$^2$.
\item{} Fig. 3 - Photons dressing the quark-gluon vertex during the 
production
 of sea quarks. The probability of producing a sea quark pair will depend 
 on 
the charge of the created quarks(Fig. 3b) and on the quark that initially 
produced the gluon(Fig. 3a).
\item{} Fig. 4 - Charge symmetry violating sea quark distributions. The 
solid, dotted, dashed, and dot-dashed curves are 
$x\delta\bar{u} =x(\bar u^p-\bar d^n)$, $x\delta \bar{d} =x(\bar d ^p-
\bar u^n)$, $x\delta s = x(s^p-s^n)$, and $x\delta\bar s=x(\bar s^p-
\bar s^n)$, respectively.
\end{itemize}
\vfill
\eject

\begin{table}
\begin{tabular}{||c|c||}\hline
Quark Content& $\delta M_\CSB$\\ \hline
$uu$ & $2(m_u-m_d)+\frac{1}{3}\epsilon_\EM =-4$ MeV\\ \hline
$ud$ & $ 0$\\ \hline
$uud$ & $(m_u-m_d)+\frac{1}{3}\epsilon_\EM =-1.3$ MeV\\ \hline
$uuud$ & $2(m_u-m_d)+\epsilon_\EM =-1.2$ MeV\\ \hline
$uudd$ & $0$ \\ \hline
$uuds$ & $(m_u-m_d)=-2.7$ MeV \\ \hline
$uud\bar{s}$ & $(m_u-m_d)+\frac{2}{3}\epsilon_\EM =0.1$ MeV \\ \hline
$uuud\bar{u}$ & $3(m_u-m_d) = -8.1$ MeV \\ \hline
$uudd\bar{d}$ & $-(m_u-m_d) +\frac{2}{3}\epsilon_\EM =5.5$ MeV\\ \hline
$uuds\bar{s}$ & $(m_u-m_d)+\frac{1}{3}\epsilon_\EM =-1.3$ MeV \\ \hline
\end{tabular}
\protect\caption{Charge Asymmetric Contributions to the mass of 
multi-quark composite
systems. $m_u-m_d=-2.7$ MeV and $\epsilon_\EM = 4.2$ MeV are determined by
fitting the neutron-proton and iso-vector diquark mass splittings.}
\end{table}

\end{document}